\documentstyle[floats,prb,aps]{revtex}

\input epsf
\begin{document}
\draft
\twocolumn[
\hsize\textwidth\columnwidth\hsize\csname@twocolumnfalse%
\endcsname

\title{Umklapp scattering from spin fluctuations in Copper-Oxides}
\author{M.\ J.\ Lercher and J.\ M.\ Wheatley}
\address{Research Centre in Superconductivity, University of Cambridge, \\
 Madingley Road, Cambridge CB3 OHE, United Kingdom.}
\date{\today}

\maketitle
\widetext
\begin{abstract}
\leftskip 54.8pt \rightskip 54.8pt

The $\mathbf k$-dependent electronic momentum relaxation rate due to Umklapp
scattering from antiferromagnetic spin fluctuations is studied within a
renormalized mean-field approach to an extended $t-J$ model appropriate to
YBa$_2$Cu$_3$O$_{7-x}$ and other cuprates.  Transport coefficients are
calculated in a relaxation time approximation.  We compare these results with
those obtained with the phenomenological assumption that all scattering
processes dissipate momentum.  We show that the latter, which violates momentum
conservation, leads to quite different magnitudes and temperature dependences
of
resistivities and Hall coefficients.

\end{abstract}
\pacs{ \leftskip 54.8pt \rightskip 54.8pt
PACS numbers: 72.10.-d,74.25.Fy,74.72.-h}
]

\narrowtext

Copper-Oxide superconductors show a systematic evolution from strong to weaker
antiferromagnetic (AFM) spin correlations as the doped hole density increases.
\cite{tranquada89} It remains unclear precisely
how these magnetic correlations are related to the intriguing transport
properties of Copper-Oxides.  Several authors
\cite{moriya90} have studied the
contribution of spin fluctuation scattering to the resistivity, assuming that
each scattering event dissipates electronic momentum.  However, it can be
argued
that cuprates are adequately described by a single band electronic
model.\cite{anderson87} In this case, the momentum transfered to the
spin fluctuation stays part of the total electronic momentum and normal
scattering does not dissipate momentum; the total electronic momentum is
conserved during the scattering event.  Umklapp processes (involving Bragg
scattering from the crystalline lattice) do not conserve total electronic
momentum and are responsible for dissipation.\cite{hellman89}

The purpose of this paper is to determine whether the distinction between
normal
and Umklapp scattering from spin fluctuations is an important one for
Copper-Oxides. For definiteness, we employ the extended $t-J$ model with
parameters appropriate to YBa$_2$Cu$_3$O$_7$ used in earlier work:
\cite{lercher94}
\begin{equation}
\label{eq:hamiltonian}
	H = -\sum_{ij\sigma} t_{ij}c_{i\sigma}^{\dag} c_{j\sigma}
	 + J \sum_{\langle ij \rangle \sigma}
	 {\mathbf S}_i \cdot {\mathbf S}_j
	 - \mu \sum_{i\sigma} c_{i\sigma}^{\dag} c_{i\sigma}
\end{equation}
where $\langle ij \rangle$ denotes a summation over nearest neighbor bonds.
$c_{i\sigma}^{\dag}$ creates an electron with spin $\sigma$ on site $i$ only if
this site is empty. $J$ is the AFM exchange constant, and the hopping integrals
$t_{ij}$ are finite for hopping between the three nearest neighbor sites.
In the auxilliary boson technique the constrained electron creation operators
are decoupled into fermionic spin and bosonic charge parts. Introducing a
uniform resonating valence bond (RVB) mean-field order parameter $v$ into the
resulting model\cite{baskaran87} leads to the fermion spectrum
$\epsilon_{\mathbf k} = - 2(\delta t + \frac{J}{2}v) (\cos{k_x}+\cos{k_y})
 - 4\delta t' \cos{k_x} \cos{k_y}
 - 4\delta t''(\cos^2{k_x}+\cos^2{k_y}-1) - \mu_{s}$,
\cite{wheatley93,lercher93} where the doping $\delta$ denotes the density of
holes and $\mu_s$ is the chemical potential for fermions. While both spin and
charge contribute to transport,\cite{ioffe89} only the contribution of the spin
part will be addressed here. Several authors have argued that the charge
(boson) part could be expected to dominate transport properties at low doped
hole density.\cite{nagaosa90}

Treating the spin-spin interaction in the random phase approximation (RPA)
leads
to the antiferromagnetically enhanced response,
\begin{equation}
\label{eq:exchange}
	\chi^{+-}({\mathbf q}, \omega) = \frac{\chi_0^{+-}({\mathbf q}, \omega)}
	{1 + J_{\mathbf q} \chi_0^{+-}({\mathbf q}, \omega)}
\end{equation}
with the Lindhard function $\chi_0^{+-}$ and $J_{\mathbf q} = J (\cos q_x +
\cos q_y)$. A mean-field commensurate AFM instability (${\mathbf Q}=(\pi,\pi)$)
occurs at low boson densities; the paramagnetic phase of interest here is
stable
at zero temperature for $\delta \gtrsim 0.15$.\cite{lercher93}

Fermion quasiparticles are strongly scattered from spin fluctuations,
especially
in the vicinity of the mean-field AFM instability in our model, where the
$(\pi,\pi)$ response grows large.  The coupling constant between fermions and
spin fluctuations is simply $J_{\mathbf q}$, which is doping independent and
large.  The quasiparticle scattering rate from $\mathbf k$ to $\mathbf{k+q}$,
denoted $\tilde{\tau}^{-1}({\mathbf{k,q}})$, is found in the inelastic Born
approximation; this is equivalent to taking twice the contribution of that
process to the imaginary part of the on-shell self energy:
\begin{eqnarray}
\label{eq:tauTilde}
	\tilde{\tau}^{-1}({\mathbf{k,q}})= &\frac{3 \hbar^2}{N}& J_{\mathbf q}^2
	\mathrm{Im}\chi^{+-}({\mathbf q},
	    \epsilon_{\mathbf{k+q}}-\epsilon_{\mathbf k})
	\\ \nonumber &\times& \left(
	\frac{1}{e^{\beta (\epsilon_{\mathbf{k+q}}-\epsilon_{\mathbf{k}})} - 1}
	+ \frac{1}{e^{\beta \epsilon_{\mathbf{k+q}}} + 1} \right) ~.
\end{eqnarray}

To relate this scattering rate to a momentum relaxation rate $\tau^{-1}
({\mathbf k})$, we assume a relaxation time picture and express the time
derivative of the {\em total} momentum as
$\case{d}{dt} {\mathbf P} = \case{d}{dt} \sum_{\mathbf k} f({\mathbf k})
{\mathbf k} = -\sum_{\mathbf k} f_0(\epsilon_{\mathbf k})
{\mathbf k} \tau^{-1}({\mathbf k})$,
where $f_0(\epsilon)$ is the Fermi function.  The change of the total momentum
can also be expressed in terms of the quasi-particle scattering rate as
$\case{d}{dt}{\mathbf P} = \sum_{\mathbf k} f_0(\epsilon_{\mathbf k})
\sum_{\mathbf q} \Delta {\mathbf{p(k,q)}} \tilde{\tau}^{-1}({\mathbf{k,q}})$,
where $\Delta {\mathbf{p(k,q)}}$ is the change in the total momentum due to the
scattering process.  Equating these two expressions gives a relation between
momentum relaxation and scattering rates,
\begin{equation}
\label{eq:requirement}
	0=\frac{1}{N}\sum_{\mathbf k} f_0(\epsilon_{\mathbf k})
	\left( {\mathbf k} \tau^{-1}({\mathbf k})
	+ \sum_{\mathbf q} \Delta {\mathbf{p(k,q)}}
	\tilde{\tau}^{-1}({\mathbf{k,q}}) \right) ~.
\end{equation}
As Equation~(\ref{eq:requirement}) was derived for the relaxation of the total
momentum, this also describes the relaxation of the total current.  We can
deduce the relaxation rate for current flowing in an arbitrary direction
${\mathbf{\hat{e}}}$; Eq.~(\ref{eq:requirement}) then reduces to
\begin{equation}
\label{eq:tau}
	\tau^{-1}({\mathbf k}) = -\sum_{\mathbf q} \tilde{\tau}^{-1}
	({\mathbf{k,q}}) \frac{\Delta \mathbf{p (k,q) \cdot \hat{e} }}
	{\mathbf{ k \cdot \hat{e} }} ~,
\end{equation}

If the momentum transfered to spin fluctuations is not considered part of the
fermion momentum, then all scattering events contribute to momentum relaxation.
Assuming ${\mathbf{\hat{e}}}={\mathbf{\hat{x}}}$, $\Delta p_x ({\mathbf{k,q}})
=
q_x$ for 'normal' processes, and $\Delta p_x ({\mathbf{k,q}}) = q_x \pm 2 \pi$
for processes with $|k_x + q_x| > \pi$.  If the momentum transfer to spin
fluctuations is part of the fermion momentum, then only Umklapp processes
dissipate momentum; $\Delta p_x ({\mathbf{k,q}}) = \pm 2 \pi$ for processes
with
$|k_x + q_x| > \pi$ and $\Delta p_x ({\mathbf{k,q}}) = 0$ for all other
processes.  Note that for the simple case of a circular Fermi surface and
$\Delta \mathbf{p(k,q) = q}$, simple symmetry arguments can be used to reduce
Equation~(\ref{eq:tau}) to the well-known result $\tau^{-1}({\mathbf k}) =
\sum_{\mathbf q} \tilde{\tau}^{-1}({\mathbf{k,q}}) (1-\cos{\Theta})$ with
the scattering angle $\Theta$.

\begin{figure}[btp]
\epsfxsize=8cm  \epsfbox{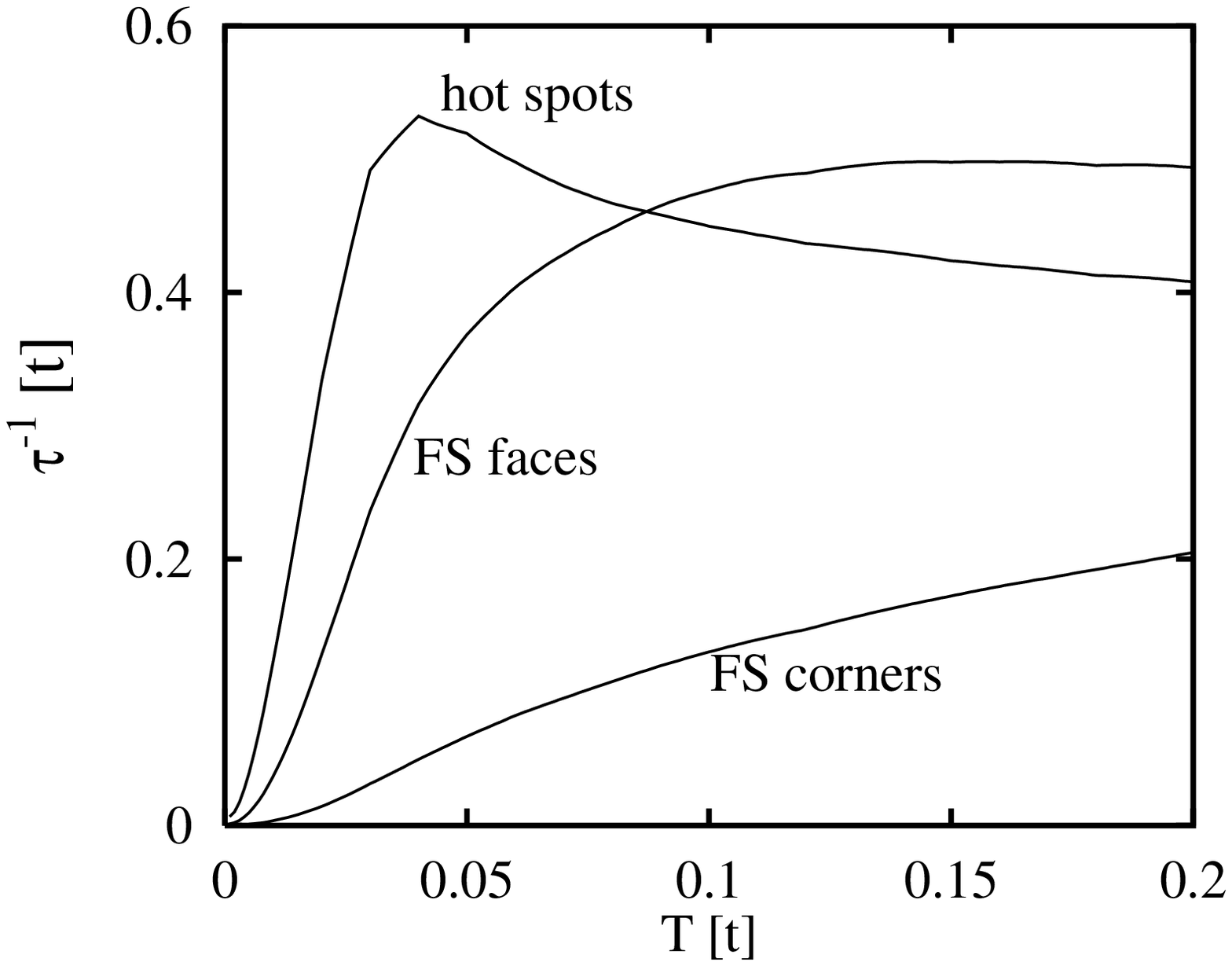}
\epsfxsize=8cm  \epsfbox{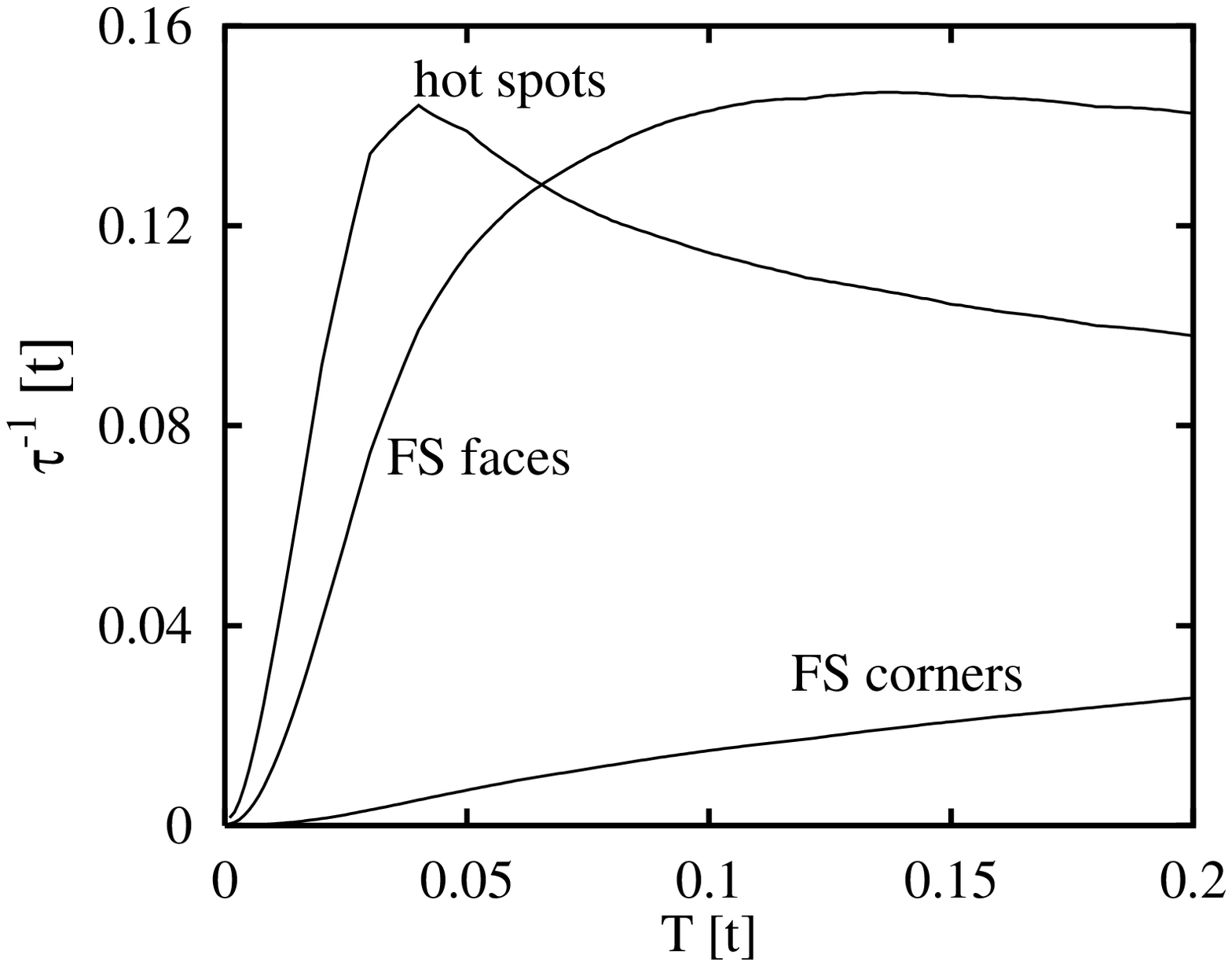}
\caption{
Temperature dependence of the momentum relaxation rate $\tau^{-1}({\mathbf k})$
at doping $\delta = 0.15$, plotted for different points $\mathbf k$ on the FS.
(a) is calculated under the phenomenological assumption that all spin
fluctuation scattering dissipates momentum, while (b) is calculated from
Umklapp
scattering alone.  The direction $\mathbf{\hat{e}}$ of the total momentum is
assumed to be parallel to $\mathbf k$ for each curve.  The different curves are
calculated at ${\mathbf k} = (0.577,\pi)$ on the FS face; ${\mathbf k} =
(1.175,1.175)$ on the FS corner; ${\mathbf k} = (0.588,2.553)$ on one of the 8
'hot spots', which are connected to other parts of the FS by the AFM ordering
vector ${\mathbf Q} = (\pi,\pi)$.  The relaxation rate is strongly anisotropic
with different $T$-dependencies on different parts of the FS.  The temperature
dependence is changed and the magnitude suppressed by a factor of $0.2-0.3$
when
only Umklapp processes are included (b).
}
\label{fig:tau}
  \end{figure}

Equations~(\ref{eq:exchange}),(\ref{eq:tauTilde}), and (\ref{eq:tau}) were used
to compute the scattering and momentum relaxation rates shown in
Figure~\ref{fig:tau} at different points on the Fermi surface (FS).  As
expected, the lifetime is found to be strongly anisotropic around the FS,
particularly in the vicinity of the AFM instability.  When only Umklapp
processes contribute (Fig.~\ref{fig:tau}(b)), this anisotropy is enhanced, but
the magnitude of the relaxation rate is reduced by a factor of $0.2-0.3$
relative to a calculation which assumes that all scattering processes dissipate
momentum (Fig.~\ref{fig:tau}(a)).  The enhancement of the anisotropy can be
easily understood from the fact that electrons on certain areas of the FS
(e.g.,
those close to the Brillouin zone (BZ) boundary) are more easily scattered
outside the BZ.  The reduction in the relaxation rate is also expected, as only
a fraction of the total scattering events can lead to Umklapp processes.  The
curves marked 'hot spots' in Fig.~\ref{fig:tau} show the momentum relaxation
rate at those points on the FS which are linked to other regions of the FS by
the AFM ordering vector ${\mathbf Q} = (\pi,\pi)$.  If $(\pi,\pi)$-scattering
were the dominant scattering process at these points, we would expect a
reduction of $\tau^{-1}$ by a factor of $\case{1}{2}$.  Instead, we find a
reduction by a factor of roughly $0.3$ at low $T$.  This indicates that even
close to an AFM instability (i.e., in a system with a long correlation length),
spin fluctuation scattering with {\em all} $\mathbf q$ contributes to the
momentum relaxation at the 'hot spots'.

\begin{figure}[btp]
\epsfxsize=8cm  \epsfbox{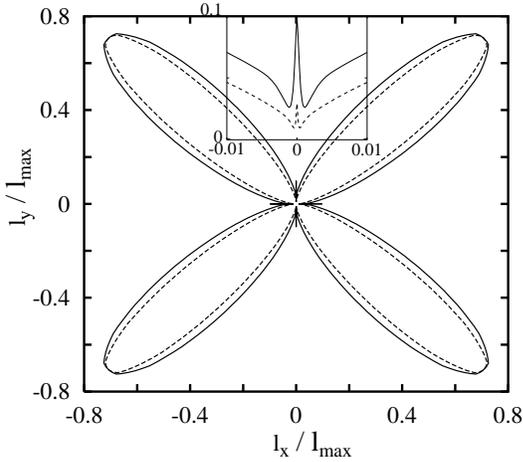}
\caption{
Anisotropy of the mean free path $\mathbf l$ around the FS at $\delta = 0.15,
T=0.015t \approx 90K$.  Each point on the curve is the endpoint of a vector
$\mathbf{l(k)}$ plotted from the origin, with $\mathbf k$ anywhere on the FS.
The curves are normalized so that $|{\mathbf l} |=1$ on the FS corners.  We
took
${\mathbf{\hat{e}}} \parallel \mathbf k$ for the calculation of each
$\mathbf{l(k)}$.  The solid line shows the mean free path calculated with all
scattering processes dissipating momentum; the dashed line is calculated from
Umklapp scattering alone.  If only Umklapp scattering dissipates momentum, the
anisotropy is enhanced.  The Hall constant is dominated by the FS corners,
where
the mean free path is long.  The effect of the hot spots (the incisions close
to
$(\pm l_x,0)$ and $(0,\pm l_y)$, see inset) on the Hall coefficient is
negligible.
}
\label{fig:MFP}
  \end{figure}

Figure~\ref{fig:MFP} is a normalized plot of the corresponding {\em transport}
mean free path (MFP) ${\mathbf{l = v}}_F ({\mathbf k}) \tau({\mathbf k})$
around
the FS.  As discussed by Ong,\cite{ong91} the 2D Hall conductance is
proportional to the area of this plot.  Thus the Hall coefficient $R_H$ is
dominated by the corners of the FS, where the spin fluctuation scattering is
weak, as discussed below.

\begin{figure}[btp]
\epsfxsize=8cm  \epsfbox{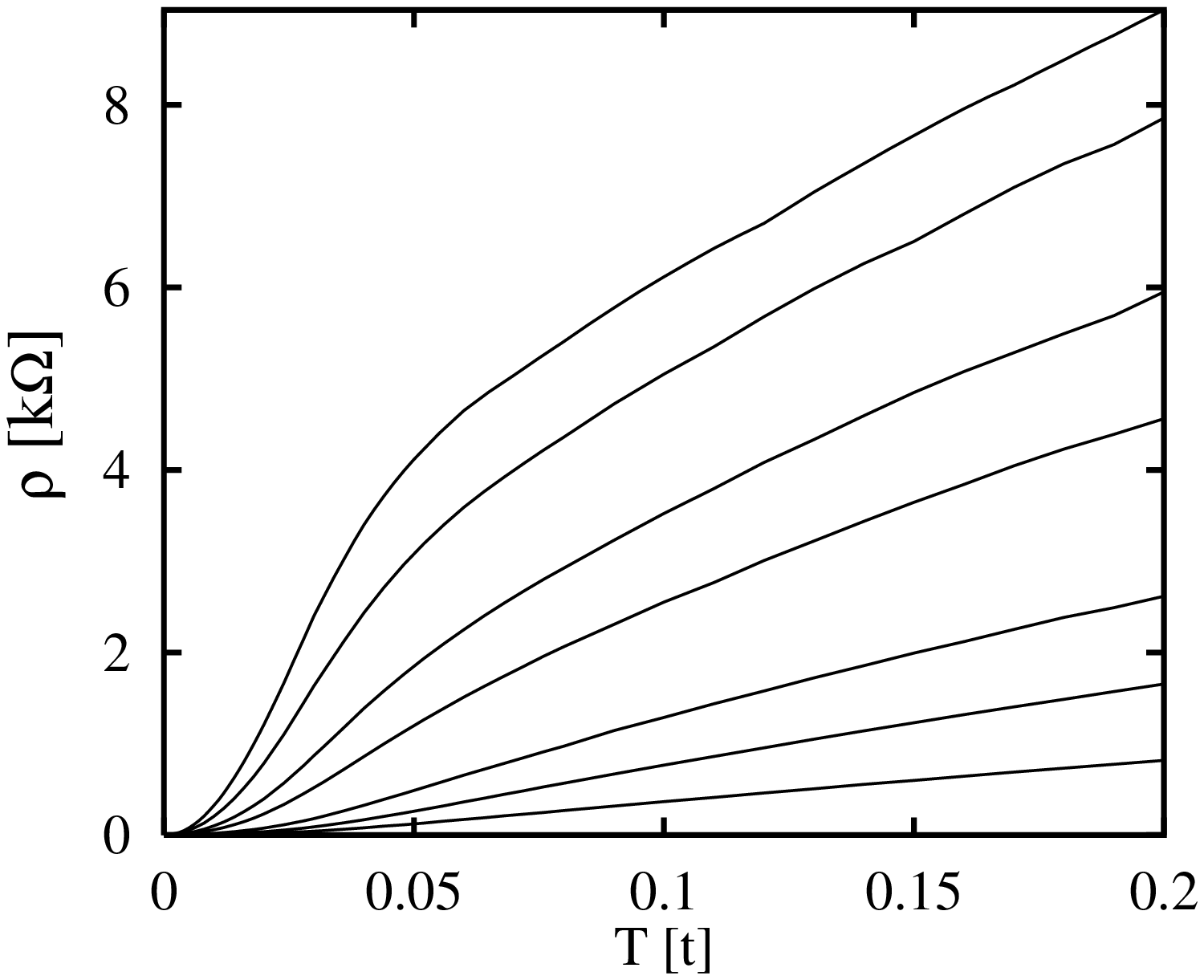}
\epsfxsize=8cm  \epsfbox{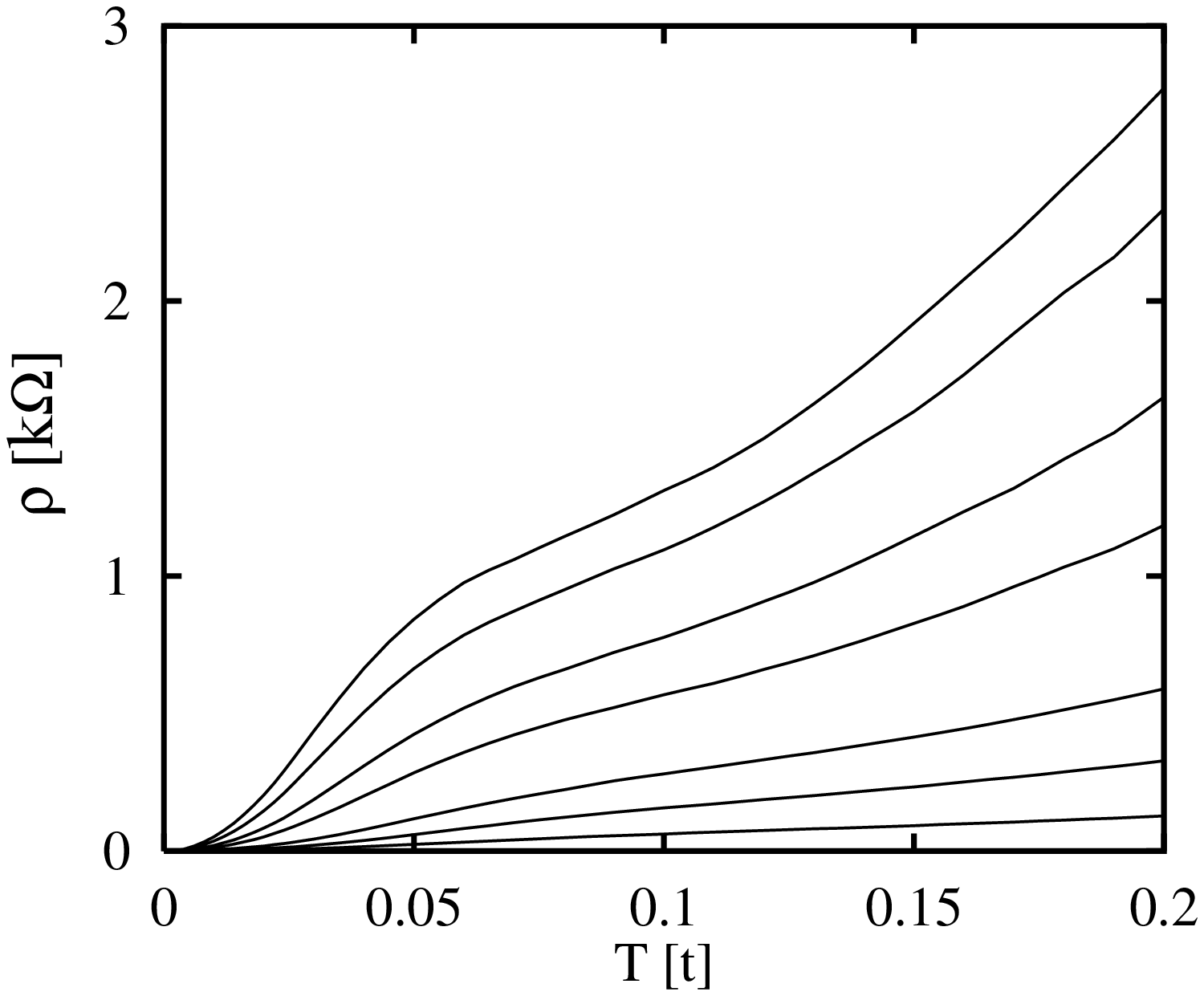}
\caption{
Temperature dependence of the resistivity per Cu atom at different doping
levels
($\delta=0.15,0.16,0.18,0.2,0.25,0.3,0.4$ with decreasing magnitude of $\rho$).
$\rho$ is calculated from all spin fluctuation scattering in (a) and from
Umklapp scattering alone in (b).  In both cases, we find $\rho \propto T^2$ at
$T \protect\lesssim T^{*} \approx 0.05t \approx 300K$.  At higher $T$, the
resistivity in (a) is linear with a positive intercept; in (b), we find $\rho
\approx A'+B'T^2$.  The magnitude of the resistivity is reduced by a factor of
$0.2-0.3$ when only Umklapp scattering dissipates momentum.
}
\label{fig:resist}
  \end{figure}

The temperature dependence of the resistivity is shown in
Figure~\ref{fig:resist}.  Irrespective of whether 'normal' processes are
included, the resistivity is found to be quadratic in $T$ at low temperatures
($T \lesssim T^* \approx 0.05t \approx 300K$).  The temperature dependence at
high $T$ is very different however.  Normal processes are responsible for the
linear resistivity $\rho \approx A + B T$ with $A>0$ shown in
Figure~\ref{fig:resist}(a).  The high temperature Umklapp scattering results
can
be described by $\rho \approx A' + B'T^2$ (Figure~\ref{fig:resist}(b)).  The
latter is smaller by a factor of $0.2-0.3$ compared to the
former.  It should be remembered that our model overestimates the magnitude of
spin fluctuation scattering in Copper-Oxides.  Nevertheless, it is worth
remarking that while the magnitude of the resistivity calculated including
normal processes at $\delta = 0.15$ is comparable to experimental results on
YBa$_2$Cu$_3$O$_7$ ($\rho_a =2.5 k \Omega $ at $T=275K \approx 0.045t$)
\cite{friedmann90}, this correspondence is certainly lost when we restrict to
Umklapp processes.

\begin{figure}[btp]
\epsfxsize=8cm  \epsfbox{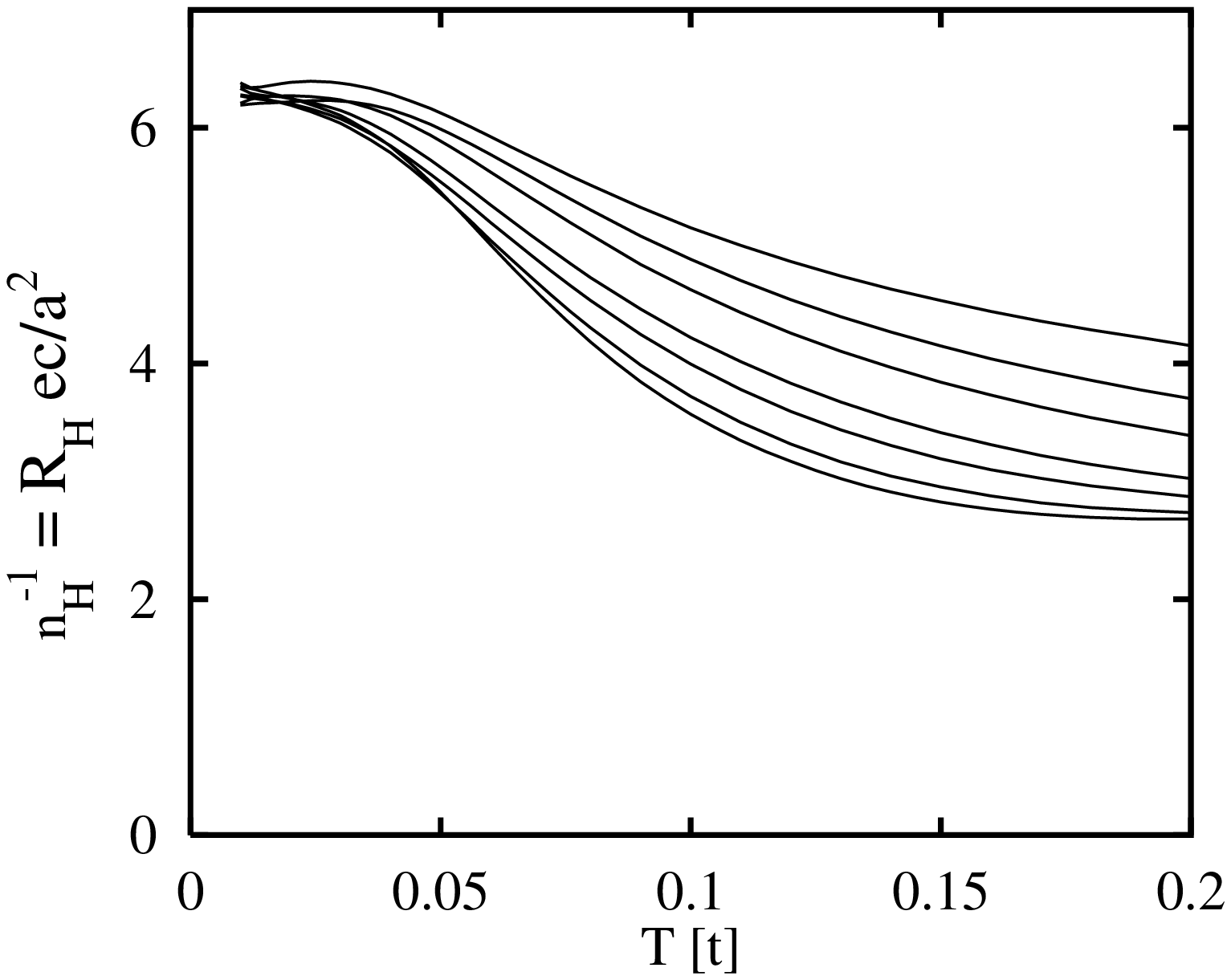}
\epsfxsize=8cm  \epsfbox{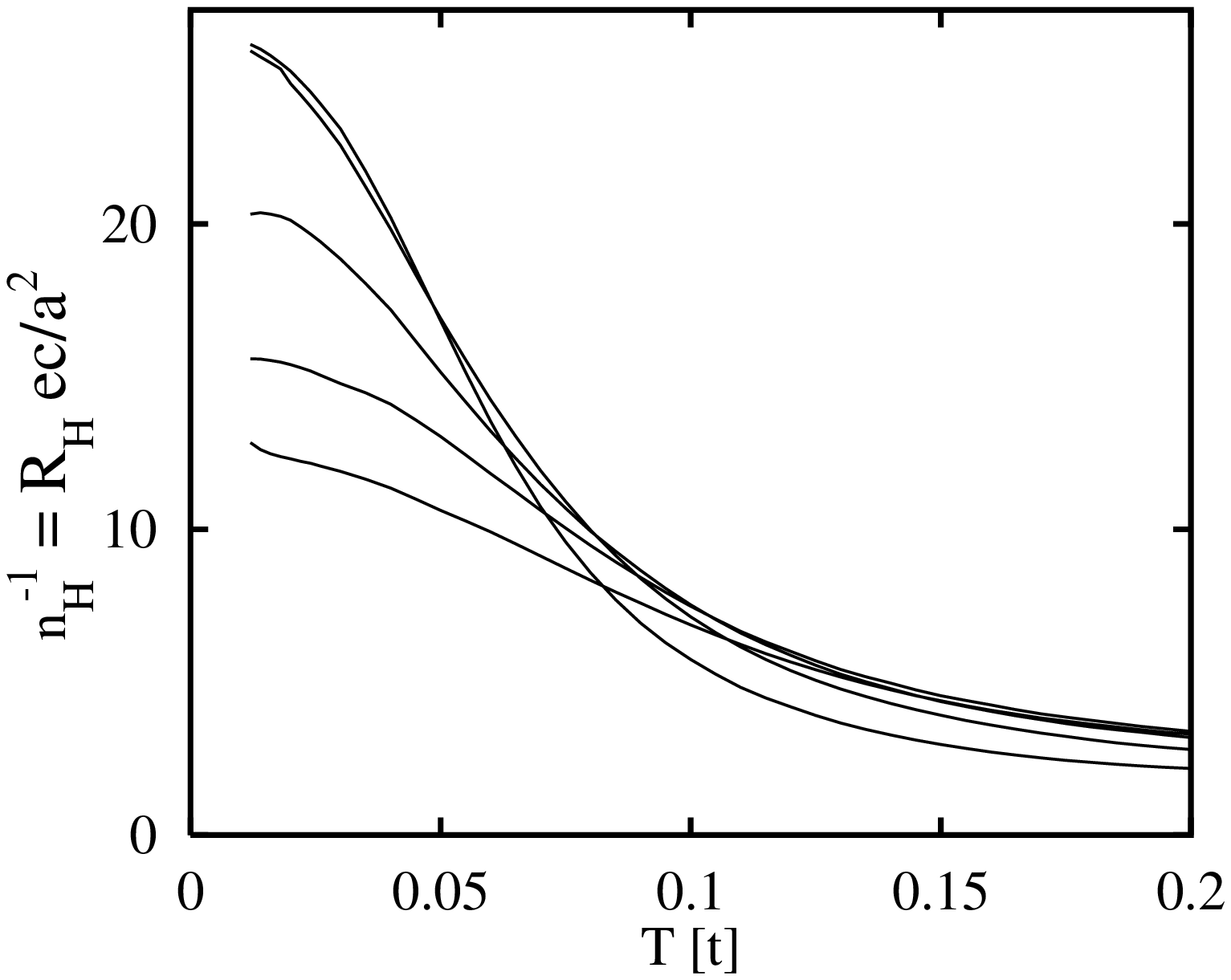}
\caption{
Temperature dependence of the Hall coefficient $n_H^{-1} = \case{ec}{a^2} R_H$
at different doping levels ($\delta=0.15,0.2,0.25,0.3,0.4$ with increasing
magnitude of $R_H$ at high $T$ in (a), and with decreasing magnitude at low $T$
in (b)).  (a) was calculated using all spin fluctuation scattering, (b) was
calculated from Umklapp scattering alone.  The temperature and doping
dependence
and the magnitude of $R_H$ are enhanced when only Umklapp scattering
contributes
(b).
}
\label{fig:R_H}

\end{figure}

\begin{figure}[btp]

\epsfxsize=8cm \epsfbox{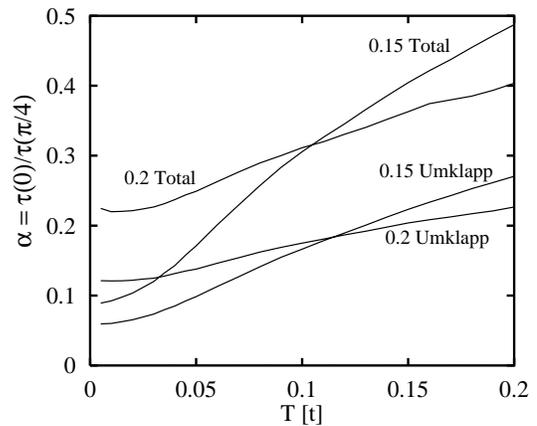}
\caption{
Temperature dependence of the anisotropy coefficient $\alpha =
\tau^{-1}(\case{\pi}{4}) /\tau^{-1}(0)$.  $\alpha=0$ corresponds to the
extremely anisotropic, $\alpha=1$ to the isotropic case.  $\alpha$ is plotted
for $\delta=0.15$ and $\delta=0.2$, assuming that all scattering events
dissipate momentum ('Total') or including Umklapp scattering only.  The
anisotropy is constant at small and very large $T$, and approximately linear in
$T$ at intermediate temperatures.
}
\label{fig:anisotropy}
  \end{figure}

Figure~\ref{fig:R_H} shows the Hall coefficient $R_H = \frac{a^2}{e c n_H}$.
Despite the single band nature of the model, $R_H$ has a strong temperature
dependence due to the anisotropy of the scattering rate around the FS.  The
$T$-dependence can be understood qualitatively using a simple parametrization
of
the momentum relaxation rate.  For each point $\mathbf k$ on the FS, we define
$\theta$ as the angle between $\mathbf{k-Q}$ and the $x$-axis, with ${\mathbf
Q}
= (\pi,\pi)$ being the center of the hole-like FS.  We parametrize the momentum
relaxation rate $\tau^{-1}(\theta)$ in terms of its average around the FS,
$\tau_0^{-1} = \langle \tau^{-1}(\theta) \rangle_{FS}$, and the ratio of its
values on the FS corner and face, $\alpha =\tau^{-1}(\case{\pi}{4})/
\tau^{-1}(0)$, as $\tau^{-1}(\theta) = \tau_0^{-1} (1+\frac{1-\alpha}{1+\alpha}
\cos{4\theta})$.  Assuming a circular FS with an isotropic Fermi velocity, this
leads to the Hall coefficient $R_H = R_0 \frac{1+\alpha}{2 \sqrt{\alpha}}$,
where $R_0 = \frac{2 \pi}{ |e| c k_F^2}$ is the $T$-independent Hall
coefficient
of the isotropic model ($\alpha=1$).  We have also included the anisotropy of
the FS and Fermi velocity through a corresponding parametrization; this gives
corrections to $R_H$ which are smaller than 3\% for the present model.
Figure~\ref{fig:anisotropy} shows the $T$-dependence of the anisotropy
parameter
$\alpha$ for our model.  $\alpha$ is found to be independent of $T$ for
$T\lesssim T_0 \approx 0.03t$ and asymptotically approaches 1 for very large
$T$
($T \gg 0.2t \approx 1200K$).  Between these regions, the anisotropy parameter
is approximately linear in $T$:  $\alpha \approx \alpha_0 + CT$.  Observing
that
$R_H$ diverges for $\alpha \rightarrow 0$ and approaches $R_0$ for $\alpha
\rightarrow 1$, the temperature dependence of $\alpha$ results in an Hall
coefficient that is constant at low and high $T$, and falls off as $T^{-1/2}$
in
the region of small to intermediate $T$.  Notice that the Hall coefficient is
strongly enhanced relative to $R_0 = \frac{2 \pi}{|e| c k_F^2}$.

In conclusion, we have argued that the microscopic origin of the spin
fluctuation spectrum determines whether or not normal processes contribute to
momentum relaxation of fermion quasi-particles.  In particular, for single band
models, only Umklapp processes contribute.  We have shown that this distinction
is a qualitatively significant one for Copper-Oxides.  In particular, spin
fluctuation scattering of fermion quasi-particles is dramatically weakened
relative to a phenomenological assumption in which normal scattering
contributes
to momentum relaxation.  In our model, the resulting resistivity is
significantly smaller than observed sheet resistances of Copper-Oxides, despite
the fact that the model overestimates the AFM spin correlations.  The fermion
Hall coefficient, on the other hand, is dramatically enhanced relative to
$\frac{2 \pi}{ |e| c k_F^2}$, growing as $T^{-1/2}$ at intermediate
temperatures.

M.J.L.\  was supported by the German Academic Exchange Service and the Federal
Ministry for Research and Development.


\end{document}